\documentstyle[prd,aps,epsfig]{revtex}

\begin{document}


\title{Effects of nuclear re-interactions in 
       quasi elastic neutrino-nucleus scattering}
\author{C. Bleve, G. C\`o, I. De Mitri, P. Bernardini,
        G. Mancarella, D. Martello, and A. Surdo      }
\address{Dipartimento di Fisica, 
         Universit\`a di Lecce and INFN, 73100 Lecce - Italy}

\date{\today}
\maketitle

\begin{abstract}
The effects of nuclear re-interactions in the quasi elastic 
neutrino-nucleus scattering are investigated with a 
phenomenological model. 
We found that the nuclear responses are lowered and that their maxima are
shifted towards higher excitation energies.
This is reflected on the total $\nu$-nucleus cross section in a
general reduction of about $15\%$ for neutrino energies above 300 MeV.
\end{abstract}



\pretolerance=100
\normalsize

\section{Introduction}              

Many works on the measurements of atmospheric and solar neutrino fluxes 
pointed out several anomalies which could hardly be explained without 
assuming the phenomenon of flavour oscillation 
\cite{pon57,zub98}.   
In particular, the flux of solar $\nu$'s is shown to be depleted in the whole
energy spectrum \cite{zub98,ber99}, while anomalies in the angular 
distributions and event rates come out from different experiments which 
measure the fluxes of $\nu$'s produced by $\pi$, $K$ and $\mu$ decays in 
the cascades initiated by primary cosmic rays in the atmosphere 
\cite{fuk98,amb98,all99}.
The continuously growing interest in the physics of neutrino 
oscillations drove an increasing effort in the studies of the various 
aspects related to the aforementioned measurements.
The correct analysis of the data
requires a good knowledge of the $\nu$--nucleon cross
section in a wide range of energies. 

The total cross section is usually calculated by summing three
separated contributions
each of them evaluated using different models
which take into account the dominant physical effects in the
various energy regions.
At energies above 3 GeV the cross section is well described 
by deep inelastic scattering processes within the parton model.
The intermediate energy region, from 1 to 3 GeV, 
is dominated by the nucleon resonances.
In this region,
the appropriate degrees of freedom to describe the cross section 
seem to be nucleons and mesons.
At energies below 1 GeV it is possible to neglect the nucleon 
excitations but it is necessary to consider nuclear effects.

A deep knowledge of the cross section for $\nu$-induced 
reactions on nuclear targets is needed for the correct analysis and
interpretation of the experimental data \cite{cav97,bat98}. 
The total inclusive cross section and, therefore, the expected number
of events could change. 
Moreover the kinematics of the final state is modified by nuclear recoil, 
Fermi motion and re-scattering. Therefore the knowledge of these effects 
provides the ultimate limit to some experimental issues.
One of these is the reconstruction 
of the direction of the incoming neutrino,
and then its pathlength from the production point 
\protect\cite{3dflux}. 
Another is the background rejection based on 
parameters like the missing momentum in the 
transverse plane, used for $\nu_\tau$ detection/tagging on long/short 
baseline experiments \cite{ast99}.

Recently, the scattering off nuclei like $^{12}$C, $^{16}$O 
(i.e. the main constituents of scintillator 
and water \u{C}erenkov detectors) has been the object of many
investigations \cite{mar99,vog98}.

In this work we shall concentrate on the 
quasi elastic regime of $\nu$ charged current interactions
\begin{equation}
  \label{qel}
    \nu_{l} + n \rightarrow l^- + p 
    ~~~~~~~~~~\mbox {and} ~~~~~~~~~~
    \bar {\nu}_{l} + p \rightarrow l^+ + n .
\end{equation}
The interest in these particular channels lies on two different aspects:
\begin{itemize}
\item The quasi elastic scattering is the dominant process at
$\nu$ energy below 1 GeV.
In long baseline or atmospheric $\nu$ experiments, this energy region is 
sensitive to low values of $\Delta$m$^2$ (i.e. $10^{-3} \div 10^{-2} \, eV^2$), 
which is indicated as the preferred 
solution for atmospheric $\nu$ anomaly.
\item The kinematics of the final state can be easily reconstructed in several 
types of detectors and provides a very clean signal if compared to the
case of  
deep inelastic scattering which produces showering events.
\end{itemize}

The quasi elastic regime is characterized by nuclear excitation
energies, i.e. the energy lost by $\nu$, whose values 
range from 30 MeV up to about 300 MeV. 
In this energy region the scattering is dominated by the 
direct interaction of the $\nu$ with a single nucleon, i.e. the
elementary processes of (\ref{qel}), while the
other nucleons act as spectators. 
Anyway the $\nu$-nucleon cross section has to be corrected because the
nucleon is not free but embedded in the nucleus.

At a first sight, the quasi elastic cross section seems
to be well described by mean--field (MF)  models, like Hartee--Fock,
shell model and Fermi gas model.
These models neglect the collective nuclear excitations which are
important at excitation energies smaller than 30 MeV, but they are not
present at higher energies.
In MF models the electro-weak excitation of the full nuclear system 
is described as the
transition of a nucleon from a state below to one above
the Fermi surface. 
In this way the binding of the nucleon in the nucleus
and the Pauli blocking are taken into account. 
A 20 years experience in the study of quasi elastic electromagnetic
excitations has shown that this picture is unable to provide 
an accurate  quantitative description of experimental data. 
The agreement with the experiment is obtained
if the rescattering between the emitted nucleon and the rest nucleus is
considered in addition to the MF effects \cite{ama99}. 

In this paper we investigate the effects of the nucleon
re-interaction on the quasi elastic $\nu$ cross sections for
all the types of neutrinos, using the phenomenological folding model
developed in refs. \cite{smi88,co88}. 
In sect. \ref{remodel} we present the folding model and we apply it  
to the nuclear Fermi gas model in sect. \ref{speapl}.
The results of the calculations are shown and discussed in
sect. \ref{resdis} and conclusions are presented in
sect. \ref{conclu}. Since in its original formulation the 
folding model was constructed to correct response functions, 
we recall in the Appendix  the relationships between
response functions, Green's functions and cross sections.

\section{Re-interaction effects}
\label{remodel}

A general analysis of charge--changing semi--leptonic weak interaction
in nuclei can be done in close analogy with electron scattering off
nuclei \cite{wal75}. 
The hypotheses usually done in the derivation of the
electron scattering cross section can also be applied 
to the case of the scattering of weak interacting probes.
A first hypothesis consists in assuming that the process can be 
well described already at the  first order Feynmann diagram i.e.
by considering only those diagrams where a single gauge boson is
exchanged. In a second hypothesis the exchanged boson is assumed to
interact with a single nucleon in the nuclear interior. 
The nuclear transition amplitude is obtained by summing the transition 
amplitudes of the single nucleons. In the second quantization language
this means that we consider  one--body transition operators only.
A further hypothesis consists in assuming that the nucleus makes a
transition between states of definite angular momentum.

In the nuclear MF model, these states are Slater determinants of the
single particle states. Since
we have chosen to restrict ourselves to the case where there is only
one-particle in the continuum the nuclear final state $|f >$
will be described
by a pure one--particle one--hole (1p--1h) excitation: 
\begin{equation} 
|f > \rightarrow | \Phi_f> = a_{p}^+ a_h |\Phi_i>, 
\label{sla1p1h}
\end{equation}
where $|\Phi_f>$ and $|\Phi_i>$ indicate the Slater determinants
describing the initial and final MF states. 

As already mentioned in the introduction,
the application of the MF model to the description of quasi elastic
electron scattering data is unsatisfactory. The great experimental
work done in this field has allowed the separation of the two response
functions, the charge and the current responses  forming the inclusive 
electron scattering cross section (Rosenbluth separation). 
MF calculations overestimate the charge responses and underestimate
the current ones (for a review see for example \cite{ama99}).

The studies done to clarify this puzzle have shown that the main
correction to the MF responses is coming from
the re--interaction of the emitted nucleon with the rest of the
nucleus. 
This effect is only partially taken into account by the Random Phase
Approximation (RPA), a theory which describes collective
excited states as linear combination of 1p-1h and 1h-1p excitations of
the ground state. 
In the quasi elastic region, i.e. at nuclear excitation energies 
above $\sim$100 MeV, continuum RPA calculations done with a finite range
effective nucleon--nucleon interaction produce responses which do not
differ very much from the MF responses \cite{co88,bub91,eng93}.
More important are those re--interaction effects beyond the RPA
description which, in the electron scattering literature, are called
Final State Interactions (FSI). The FSI take into account the
possibility that, after the interaction with the probe, the
nucleus remains in a highly excited state which can be described only
in terms of many--particle many--hole excitations. 
In this case FSI lower by $15 \div 20\%$ the MF responses.

We think that FSI may play an important role also in the  description of
weak quasi elastic responses. 
We have considered FSI using the model developed in \cite {smi88,co88}, 
which we shall briefly present in the following.

The full Hilbert space $\cal H$ can be divided 
in a subspace ${\cal H}_0$  composed by all the 
1p-1h Slater determinants defined in eq. (\ref{sla1p1h}) and a
complementary subspace ${\cal H}_c ={\cal H } - {\cal H}_0$.
The 1p-1h excited states are eigenvectors of the MF hamiltonian $H_0$ 
\begin{equation}
H_0 |\Phi_f> = E^{MF}_f |\Phi_f> \,\,,
\end{equation}
and lead to the MF expression of the response:
\begin{eqnarray}
S^{MF} (|{\bf q}| , \omega) 
&=& \sum_f <\Phi_f|{\cal O}({\bf q})|\Phi_i> 
         <\Phi_f|{\cal O}({\bf q})|\Phi_i>^\dagger
         \delta(E_f-\omega) \\
& =& \displaystyle{-\frac{1}{\pi}} 
Im <\Phi_i| {\cal O}^+({\bf q})   
G^0( \omega ) 
{\cal O}({\bf q})|\Phi_i> \,\,.
\label{resmf}
\end{eqnarray}
where we have indicated with $G^0( \omega )$ the MF Green's function
(see Appendix). 

To evaluate the modifications of $S^{MF}$ due to presence of  
${\cal H}_c$ we introduce projection operators $P$ onto ${\cal H}_0$ 
and $Q$ onto ${\cal H}_c$. The effects of ${\cal H}_c$ can be
taken into account by considering the effective hamiltonian:
\begin{equation} 
H_{eff}(\omega) = PHP - PHQ \,\,
\displaystyle{\frac{1}{QHQ - \omega - i \eta}} \,\,QHP
= PHP - U(\omega) \,\,.
\end{equation}

If the hamiltonian $H_0$ is composed by one-body hamiltonians, like in
the MF case, only the two--body term $V=H-H_0$ can connect
the ${\cal H}_0$ and ${\cal H}_c$ Hilbert subspaces, therefore the
many-particle many--hole excited states forming  ${\cal H}_c$ act only
as doorway states. Since the operators ${\cal O}({\bf q})$ are
one--body operators they act only on the ${\cal H}_0$ subspace.

Using the effective hamiltonian in the Green's function 
( eq. (\ref{fullgreen}) in the appendix)
and inserting a full set of eigenstates of $H_0$, 
we obtain the following expression for the response function:
\begin{eqnarray} 
S (|{\bf q}| , \omega) & = & \displaystyle{\frac{1}{\pi}}  
 Im \sum\limits_{f,f'} 
 < i| {\cal O}^+({\bf q}) |\Phi_f> \, \left[
 \displaystyle{\frac{1}{E_f - \omega - \Sigma_{f f'}(\omega) - i \eta}
 } \right. \nonumber \\
&~& + 
\left.  \frac{1}{E_f + \omega - \Sigma_{f f'}(- \omega) + i \eta} \, \right]
<\Phi_{f'} | {\cal O}({\bf q})|i> \,\, ,
\label{reseff}
\end{eqnarray}
where we have used 
$< \Phi_f | PHP|\Phi_{f'}> =< \Phi_f | H_0|\Phi_{f'}> = E_f \delta_{ff'}$ 
and we have defined
$\Sigma_{f f'}( \omega )=< \Phi_f | U(\omega) |\Phi_{f'}>$.

We introduce the assumption 
that in quasi elastic energy range the term $\Sigma_{f f'}$ 
does not strongly depend on the individual MF states $|\Phi_f>$ but
rather on the full phase space available at the energy $\omega$.
In other words we make the assumption: 
\begin{equation}
\Sigma_{f f'}( \omega ) \longrightarrow \Sigma (\omega) \delta_{ff'} \,\, ,
\end{equation}
where
$ \Sigma(\omega)$ is a complex function of the transferred energy:
\begin{equation}
\Sigma (\omega) = \Delta (\omega) - \frac{i}{2} \Gamma(\omega) \,\,.
\label{sigma}
\end{equation}
Using the above expressions in eq.(\ref{reseff}) 
we obtain:
\begin{equation}
S (|{\bf q}| , \omega) = 
\sum\limits_{f }
 |< \Phi_f | {\cal O}({\bf q}) |i>|^2
 \left[ \rho (E_{f}, \omega) + \rho (E_{f}, - \omega) \right] \,\, ,
\label{resfin}
\end{equation}
where we have defined:
\begin{equation}
\rho (E, \omega) = 
\displaystyle{\frac{1}{2 \pi} \frac{\Gamma (\omega)}
{ \left[ E - \omega - \Delta (\omega) \right]^2 + 
  \left[ \Gamma (\omega) /2 \right]^2}} \,\, .
\label{confun}
\end{equation}

In the expression (\ref{resfin}) the nuclear ground state is still
eigenstate of the full hamiltonian $H$, but it is usually replaced by the
nuclear ground state provided by $H_0$. Making this substitution and
considering that the quasi elastic region is in the continuum
excitation region of the nucleus, we can rewrite the 
full response function as a convolution of the MF response
(\ref{resmf}) with the lorentzian functions $\rho(E,\omega)$:
\begin{equation}
S (|{\bf q}| , \omega) = 
\int_0^\infty dE \, S^{MF} (|{\bf q}| , E)
\left[ \rho (E, \omega) + \rho (E, - \omega) \right] \,\, 
\label{rescon}
\end{equation}

The FSI produce three effects on the MF
response: a lowering of the maximum value of the
response, a widening of the width and a shift of the position of the
peak due to the $\Delta (\omega)$ term.

The functions $\Delta (\omega)$ and $\Gamma (\omega)$ 
are connected by a dispersion relation:
\begin{equation}
\Delta (\omega) = \frac{1}{2 \pi} P \int_{- \infty}^{+ \infty}
 d \omega ' \frac{\Gamma(\omega ')}{\omega ' - \omega} \,\, ,
\end{equation}
where we have indicated with $P$ the principal value integral.
Therefore our calculations would only require the knowledge
of $\Gamma( \omega)$ which 
can be related to the imaginary part of the single particle
self-energy.

Our description of the nuclear excitation 
does not consider 
the non locality of the mean
field. 
A simple way to take  into account this important correction
 is the introduction of a
$q$--dependent nucleon effective mass $M^*(q)$ \cite{mah82}.
For a given value of $M^*/M$ the following scaling relation 
holds:
\begin{equation}
  S_{M^*} (|{\bf q}| , \omega) = \frac{M^*}{M} S_{M} (|{\bf q}| , 
  \frac{M^*}{M} \omega) \,\,.
\label{resmef}
\end{equation}

\section{Application to the Fermi gas model}
\label{speapl}
In the previous outline of the folding model 
we have not specified the characteristics of the MF hamiltonian $H_0$ 
(the only requirement was that it
should be a sum of single particle hamiltonians).

Finite nuclear systems are realistically described by MF hamiltonians
of Hartree--Fock and shell model type where the single particle
wave functions are calculated solving the single particle
Schr\"o\-din\-ger 
equation for a spherically symmetric potential. 
On the other hand, the excitation energy and 3-momentum transfer
values characterizing the quasi elastic region are such that surface
and collective excitations of the nucleus are not important. 
Therefore it is plausible to use in this region a simpler 
MF model, the Fermi gas (FG) model, which describes the nucleus as a 
translationally invariant system composed by an infinite number of nucleons
whose momentum distribution is given by:
\begin{equation}
n(|{\bf p}|) = \displaystyle{\frac{{\cal T}}{\frac{4}{3} \pi k^3_F}} \Theta
        (k_F - |{\bf p}|) \,\, ,
\end{equation}
where $\Theta$ is the Heaviside function,
${\cal T}$ = Z or N, $k_F$ is the Fermi momentum and ${\bf p}$ the
nucleon 3-momentum.

In this system the full hamiltonian is the sum of free single particle
hamiltonians and the single particle wave functions used to form
Slater determinants are plane waves. 
The nucleon single particle energies $\epsilon (|{\bf p}|)$ 
are related to their 
3-momentum by
\begin{equation}
       \epsilon (|{\bf p}|)  = \sqrt{ |{\bf p}|^2 + M^2} - w  \,\, ,
\end{equation}
where $p=(p_0,{\bf p})$ is the nucleon 4-momentum,  
$M$ its rest mass and we have subtracted a constant binding energy $w$.
The nucleus excitation energy is given by the
difference between the particle and hole single particle energies.
In the Fermi gas model the observable quantities are calculated per
unit of volume and per nucleon. Their values should remain constant in
the limit of infinite volume and nucleon number.

The relevance of nuclear finite size in the quasi elastic region
has been studied in \cite{ama94} by comparing electromagnetic
shell model and FG responses.
It has been found that the FG model generates nuclear
responses rather close to those of the shell model 
as long as the value
of the Fermi momentum is taken as the average value with respect to the
nuclear matter density $n(r)$:
\begin{equation}
<k_F>=\Big( \frac{3}{2} \pi^2 \Big)^{1/3}
\frac { \int dr\,r^2\,[n(r)]^{4/3} } { \int dr\,r^2 \, n(r) }
\,\, .
\label{avekf}
\end{equation}
Neutrino--nucleus quasi elastic cross section was
first evaluated in a FG model  by Smith and Moniz \cite{smi72}, and the final
expression contains the various responses (as it does the expression given in
eq. (\ref{fullcross})).
Since the interaction matrix elements show a very weak dependence on the 
initial nucleon 3-momentum ${\bf p}$, we make the common approximation of 
writing the cross section in a factorized form:
\begin{equation}
\left( \frac{d \sigma (|{\bf q}|, \omega )} {d \Omega_l d E_l} \right)_{FG} = 
\sum_{i=1}^A
\left( \frac{d \sigma (|{\bf q}|, \omega ) } {d \Omega_l}
\right)_{(\nu,N_i)} \,
R^i_{FG} (|{\bf q}|, \omega ) \,\, ,
\label{fgcross}
\end{equation}
where $A$ is the number of nucleons forming the target nucleus, ${d \Omega_l}$
is the differential solid angle around the direction of the outgoing lepton, 
and 
\begin{equation}
\left( \frac{d \sigma (|{\bf q}|, \omega ) } {d \Omega_l} \right)_{(\nu,N)}
\label{crossnun}
\end{equation}
is the neutrino cross section on free nucleons. 
In eq. (\ref{fgcross}) the nuclear effects, 
Fermi motion,  binding  and Pauli blocking,
are taken into account by the $R_{FG} ({\bf q}, \omega )$ function
whose expression is \cite{sar95}:
\begin{equation}
R^i_{FG} (|{\bf q}|, \omega ) = 
\displaystyle{ \frac{1}{\frac{4}{3} \pi k_F^3}}
\int \displaystyle{\frac{d^3p M^2_i}{p_0 p'_0}} 
        \delta (p_0 + \omega - p'_0) 
        \Theta (k_F - 
        |{\bf p}|) \Theta (|{\bf p'}| - k_F)  \,\, ,
\end{equation}
where $p=(p_0,{\bf p})$ and $p'=(p'_0,{\bf p'}={\bf p}+{\bf q})$ are
the initial and final nucleon 4-momenta.
The validity of the approximation leading to the factorized form of
the cross section has been checked in ref. \cite{sar95} where it was
found that in the quasi elastic region  the expression (\ref{fgcross}) 
numerically differs from the full expression given in ref.
\cite{smi72} by a few percent.

The folding model described in the previous section
should be  applied to the response functions using different folding
functions $\rho(E,\omega)$ for every response function  according to
their spin and isospin dependence.
In the quasi elastic region these differences among responses
are rather small, therefore we have used the same folding
function for all them. 
Since the cross section is a sum of responses, we have applied 
the folding directly to the cross section:
\begin{eqnarray}
\left( \frac{d \sigma (|{\bf q}|, \omega )} {d \Omega_l d E_l} 
\right)_{M^*} 
 &=& 
\int_0^\infty d\tilde{E} \, 
\left( \frac{d \sigma (|{\bf q}|,\tilde{E}  )} 
{d \Omega_l d E_l} \right)_{FG}
\,
\left[ \rho (\tilde{E} , \omega) + 
\rho ( \tilde{E}, - \omega) \right] 
\nonumber
\\
&=&
\sum_{i=1}^A
\left( \frac{d \sigma (|{\bf q}|, \omega ) } {d \Omega_l}
\right)_{(\nu,N_i)} \, 
R^i (|{\bf q}|, \omega ) \,
\label{fxsect}
\end{eqnarray}
following eqs.(\ref{rescon}) and (\ref{resmef}), with the definition
$\tilde{E}=(E M^*)/M$ .

\section{Results and discussion}
\label{resdis}
The only free parameter required by the FG model
is the value of the Fermi momentum which can be related to the average
density of the real nuclear system through eq. (\ref{avekf}).
In all our calculations we have used $k_F = 220 $ MeV/c which
corresponds to the average density of nuclei with mass number $ 15 < A
< 50 $.  

Once the FG cross section has been calculated, the other inputs we need
are related to correction for the re--interaction:
the $\Gamma(\omega)$ function of eq. (\ref{sigma}) and
the effective nucleon mass $M^*$.
From a theoretical point of view  $\Gamma ( \omega)$
can be evaluated by considering many--particle many--hole nuclear
excitations, however we have used an estimate based on a comparison with
experiments. 
The data we have considered for positive values of $\omega$ are those
related to the imaginary part of the nuclear mean field whose
parameters have been fixed to fit nucleon--nucleus elastic
scattering cross sections. For negative values of $\omega$ we have
considered the energy width of the single particle levels, measured in
knock--out reactions like $(e,e'p)$ or $(p,d)$. 
We have obtained $\Gamma (\omega)$
by making the average of the single particle energy width
$\gamma(\omega)$ :
\begin{equation}
\Gamma (\omega) = \frac{1}{\omega} \int^\infty_0 d \epsilon 
\left[  \gamma(\epsilon+\omega) + \gamma(\epsilon - \omega) \right] \,\, .
\end{equation}

In order to reproduce the empirical values of the single particle
widths, we have used the expression:
\begin{equation}
 \gamma ( \epsilon) = a \cdot  \frac{\epsilon^2}{\epsilon^2 + b^2}
        h(\epsilon) \,\,,
\label{spgam}
\end{equation}
with $a=11.5$ and $b=18$ \cite{mah82}. 
Unfortunately the data are limited to $|\epsilon|<100 \,$MeV, therefore 
the high energy behaviour, controlled by $h(\epsilon)$, is affected by 
strong uncertainties. We have used two extreme parameterizations compatible
with the data:
\begin{eqnarray}
   h(\epsilon) & = & 1 
   \label{he1}
   \\
   h(\epsilon) & = & \frac{c^2}{\epsilon^2+c^2}
   \label{hefun}
\end{eqnarray}
with $c=110$.

The value of the effective nucleon mass is another input of our calculation. 
It has been determined by using the expressions
proposed in the polarization potential model of ref. \cite{pin88}: 
\begin{equation}
   \frac{M^*}{M} = \frac{1}{1+2M \Delta U/(q^2+q_0^2)}
\label{emass}
\end{equation}
with 
\begin{equation}
         q_0^2 =\frac{2M \Delta U}{M/M^*_0 -1} \,\,\ , 
\end{equation}
where $\Delta U$=50 MeV is the depth of the potential well. The above
expressions have been built such as $M^*/M=1$ in the limit for 
$q \rightarrow \infty$ and $M^*=M^*_0$ for $q \rightarrow 0$.
Various nuclear structure studies \cite{mah82,pin88,vau72}
indicate that $M^*_0/M$=0.8, and this
is the value we have used in our calculations.

\begin{figure}
\begin{center}
\mbox{\epsfig{file=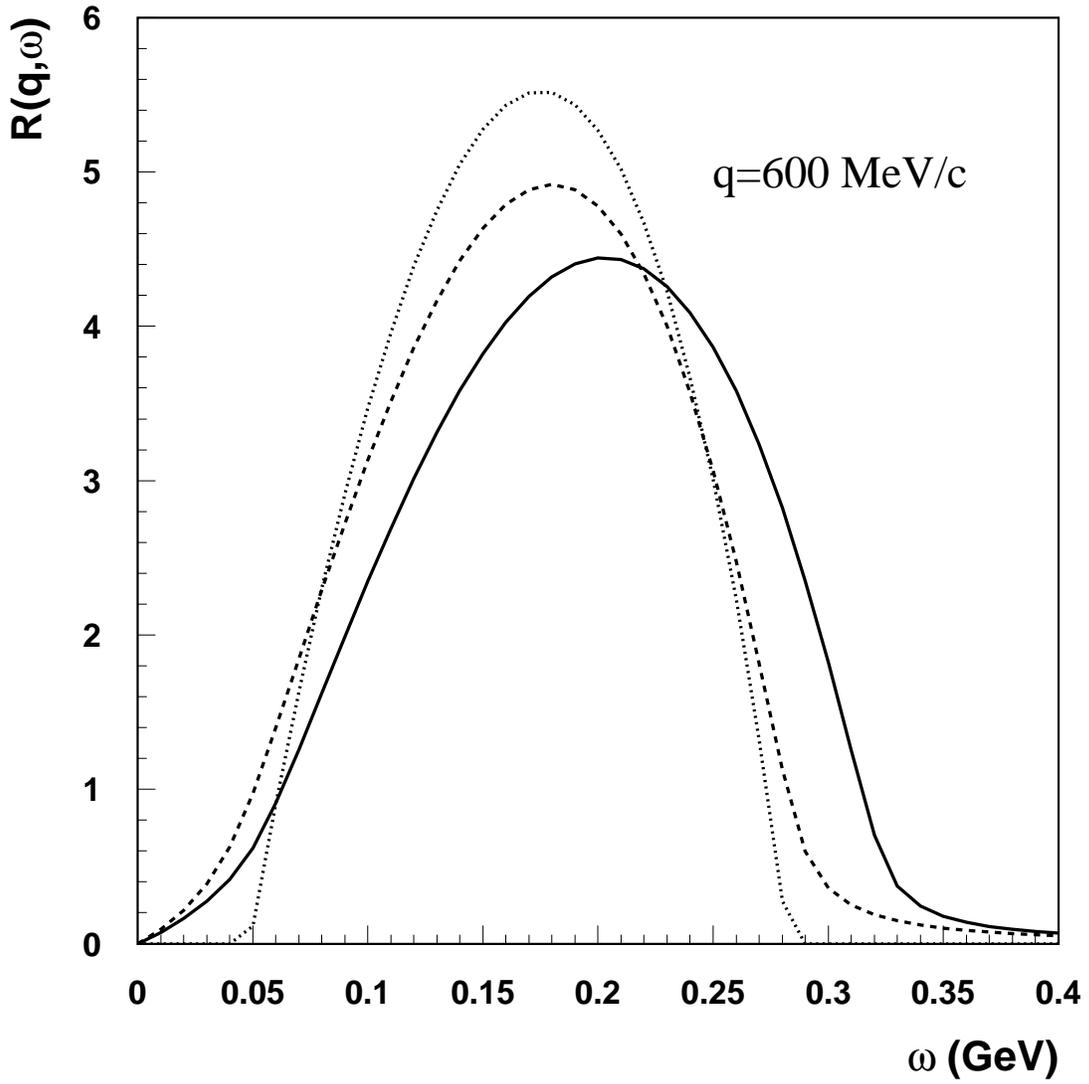,width=15cm,bbllx=31bp,bblly=200bp,bburx=550bp,bbury=630bp} }
\vskip 4.cm
\caption{Dependence of $R(|{\bf q}|,\omega)$ 
on the transferred energy $\omega$ for 
$\nu_{\mu}$ - nucleus scattering. 
The dotted line shows the result of a bare FG calculation, 
the dashed one has been obtained with $M^*=M$, and
the solid line includes all the considered FSI effects (see text).
The responses per nucleon on isoscalar target (N=Z) are represented.}
\label{figres}
\end{center}
\end{figure}

The effects of the re-scattering are evident in fig.\ref{figres} where
we show  
$R({\bf q}, \omega )$ of eq. (\ref{fxsect}) calculated for
$|{\bf q}|$=600 MeV/c as a function of $\omega$.
The dotted line represents $R_{FG}({\bf q}, \omega )$,
eq.(\ref{fgcross}). 
The other two lines have been obtained performing the folding with 
$h(\epsilon)$ given by eq. (\ref{hefun}), the dashed line with
$M^*/M=1$, and the full line using the value of the effective mass
given by the procedure discussed above.
A comparison between dashed and dotted lines shows that the 
re--interaction is moving strength from the peak region towards
both lower and higher energy regions. The total
strength is conserved i.e. the areas underlined by the two curves in the
figure are equal. 
The  peak position is slightly shifted towards higher energy values
owing to the presence of $\Delta(\omega)$ in the denominator of the folding
$\rho(E,\omega)$ functions.
The differences between dashed and full lines are due to the 
nucleon effective mass which produces a further spreading,
a lowering of the peak value and a shift of the peak to a
higher energy. 

\begin{figure}
\begin{center}
\mbox{\epsfig{file=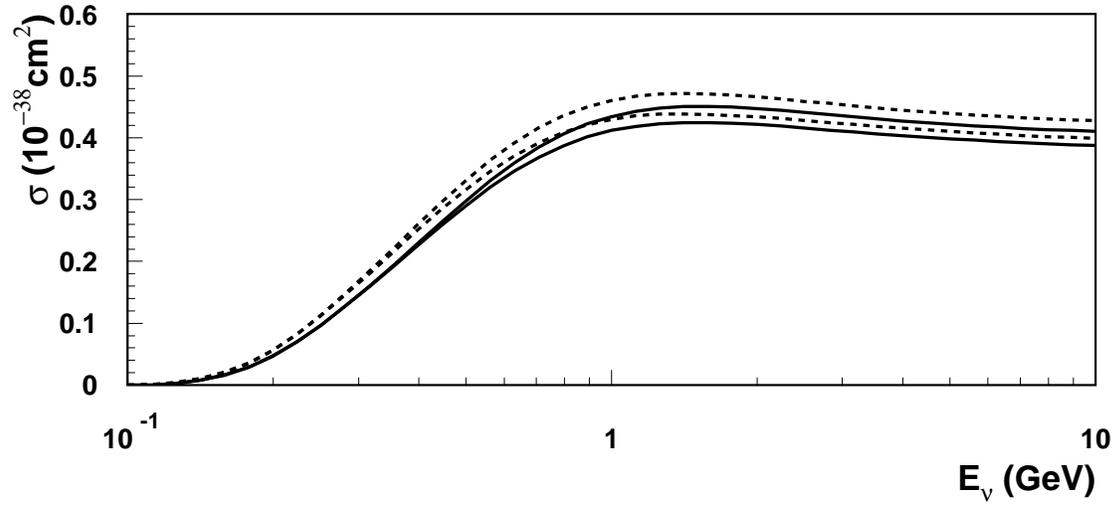,width=15cm,bbllx=31bp,bblly=200bp,bburx=550bp,bbury=630bp} }
\caption{Sensitivity of the total $\nu_{\mu}$ - nucleus cross section 
to the parametrization of $\Gamma (\omega)$ and to M$^*$/M (see text).
In this figure, and in the following ones, the cross sections per 
nucleon on isoscalar target are shown.}
\label{figunc}
\end{center}
\end{figure}

\begin{figure}
\begin{center}
\mbox{\epsfig{file=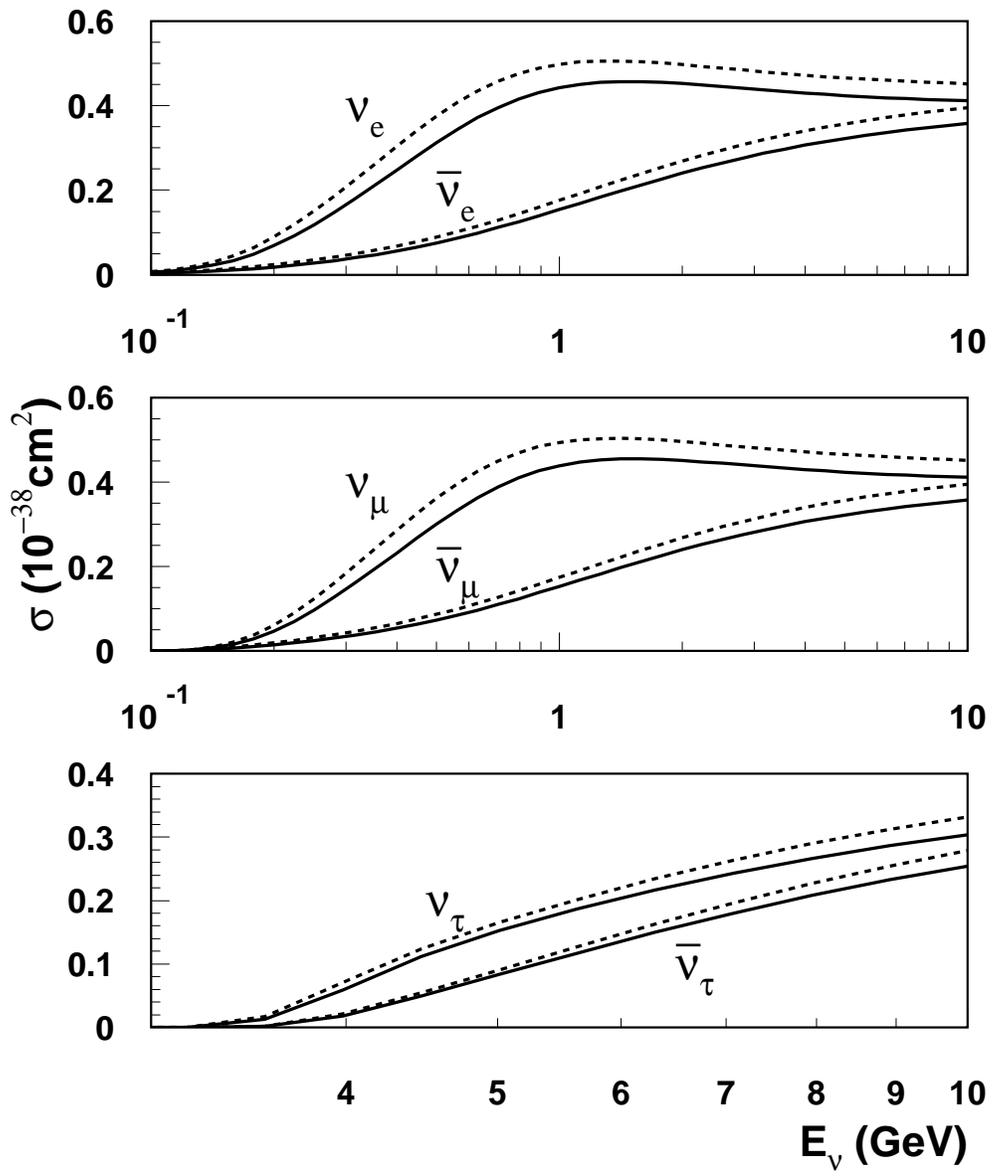,width=15cm,bbllx=31bp,bblly=200bp,bburx=550bp,bbury=630bp} }
\vskip 4.cm
\caption{Cross sections for quasi elastic $\nu$ - nucleus and 
$\bar{\nu}$ - nucleus scattering
for all flavours. The dashed lines show the result of a bare FG calculation, 
while solid lines include FSI effects.}
\label{figallmu}
\end{center}
\end{figure}

\begin{figure}
\begin{center}
\mbox{\epsfig{file=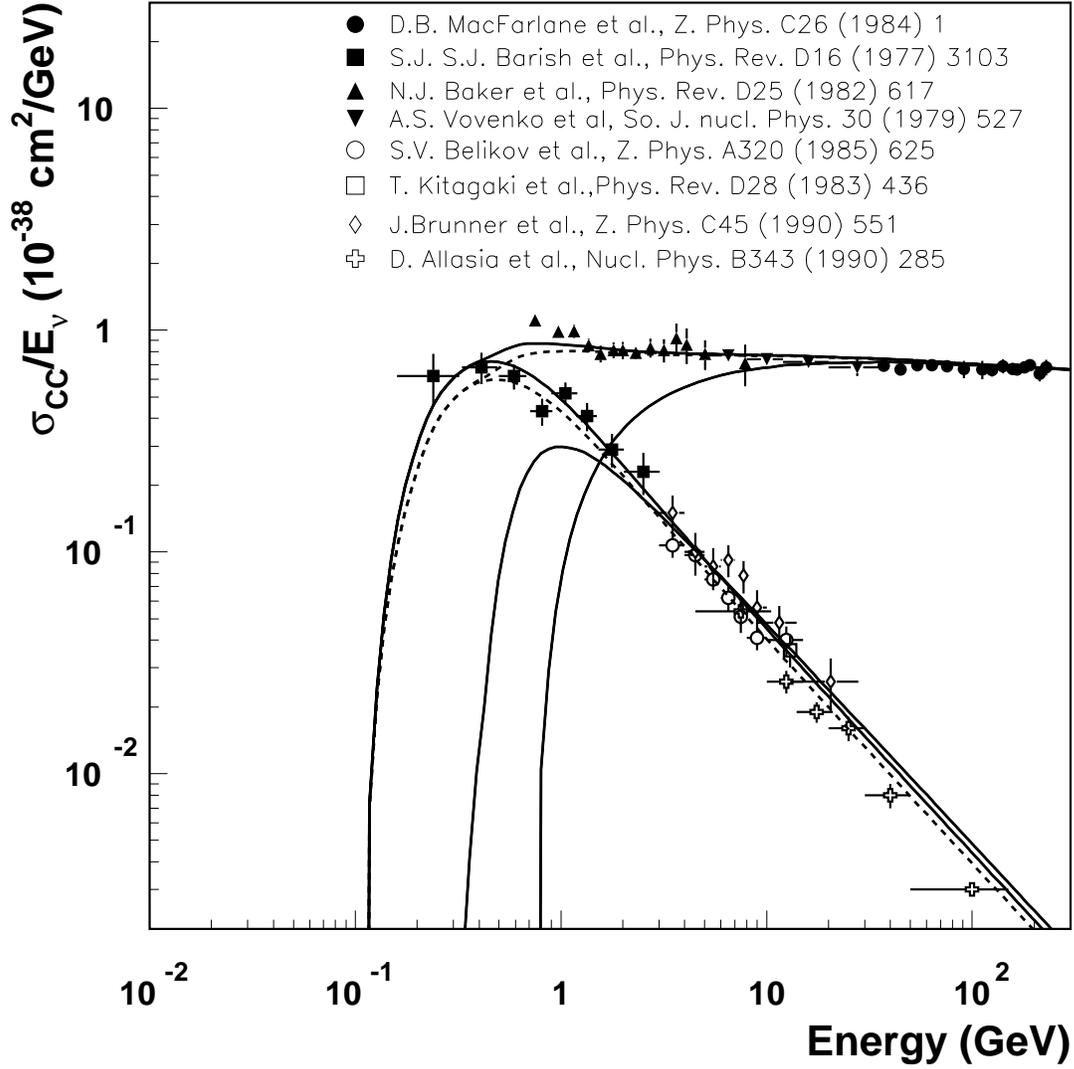,width=15.cm,bbllx=31bp,bblly=200bp,bburx=550bp,bbury=630bp} }
\vskip 4.cm
\caption{Cross section per nucleon (isoscalar target) for quasi elastic
$\nu_{\mu}$ -  nucleus scattering compared with experimental data.
The three contributions of the {\it quasi elastic}, {\it nucleon resonances} and 
{\it deep inelastic scattering}, as calculated in the framework of the FG
model \protect\cite{lip95}, are shown with solid lines. 
The dashed line shows the result of the quasi-elastic cross section 
calculated in this work including also FSI effects.
The total cross section, computed with the two different results in the quasi 
elastic region, is also reported. The $15\%$ lowering of the quasi elastic peak, 
due to Final State Interactions is clearly visible. }
\label{figexp}
\end{center}
\end{figure}

The sensitivity of our results to the expression chosen to set
$\Gamma(\omega)$ and to the effective mass, can be seen in
fig.\ref{figunc}, where we show  the 
$\nu_\mu$--nucleus cross sections per nucleon as a function of the
neutrino energy.
The dashed lines have been obtained setting $M^*/M=1$ 
while the full lines have been calculated 
using the expression (\ref{emass}). 
In both cases the upper curves have been obtained using
$h(\epsilon)=1$ in eq. (\ref{spgam}) while the lower lines 
have been produced with the expression (\ref{hefun}). 
As expected the use of the effective mass lowers the values of the
cross section at all energies. 
It is also clear that the use of $h(\epsilon)=1$ enhances the effect of
the folding, since this means that the re-scattering is active
whatever is the energy of the emitted nucleon. In the other choice the
re-scattering effects are switched off when the energy of the emitted
nucleon is above 200 MeV. 
We think that the calculations with $M^*$ given by eq.(\ref{emass})
are the most realistic ones. 
On the other hand, we do not have any good reason to prefer one of 
the two possible choices for $h(\epsilon)$.
The expression (\ref{hefun}) would be favored by  
the speculations about the so-called color transparency
\cite{ber81} claiming that the nucleus is transparent to high
energy emitted nucleons. However these speculations do not have, so
far, experimental support. 
The results presented in  figs.\ref{figallmu} and \ref{figexp} have been 
obtained using effective mass and eq. (\ref{hefun}). 
We should remark, in any case, that the maximum difference
between the upper and lower full lines in fig.\ref{figunc} is of 5\%.

\begin{figure}
\begin{center}
\mbox{\epsfig{file=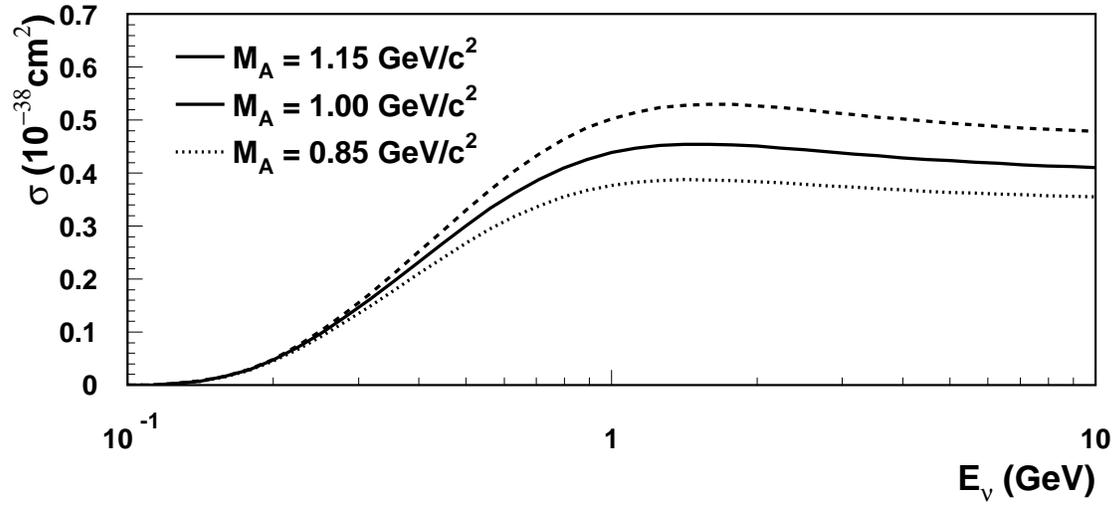,width=15cm,bbllx=31bp,bblly=200bp,bburx=550bp,bbury=630bp} }
\caption{ Sensitivity of the total $\nu_{\mu}$ - nucleus cross section
  to the value of the axial mass M$_A$ (see text). }
\label{figaxial}
\end{center}
\end{figure}

Our study has been made under the assumption of no uncertainties in 
the elementary $\nu$--nucleon cross section. 
This hypothesis, common to  all the calculations of this type, 
is not fully correct, since this elementary cross section depends
upon the electromagnetic and axial nucleon form factors.
In our calculations we have used the dipole form for the electromagnetic
form  factors, and we have checked that changes produced by
the more sophisticated parameterization of ref. \cite{hoh76} were
within the numerical accuracy of the calculations.
Since we have adopted the dipole expression,
the only free parameter is the axial mass M$_A$.
In the calculations presented so far we used the axial mass
value of $M_A=1.00$ GeV/c$^2$ which has been adopted also in
ref. \cite{lip95}. This value is compatible with the analysis of the
neutrino nucleus total cross section data of refs. \cite{kit83,bel85}. 
On the other hands, these studies, and also other ones done on
different target nuclei, show an uncertainty on the axial mass
value of ~15$\%$.  
To test the sensitivity of the cross section to the axial form factor 
we have calculated the $\nu_{\mu}$-nucleus cross section by changing the
axial mass value by a 15\%. These results are presented in
fig. \ref{figaxial} where it is shown that the changes on the
cross section are of the same order of magnitude of the changes on the
axial mass. This is not a specific feature of our model, but it is a
systematic error which affects all the theoretical evaluations of the 
$\nu$-nucleus cross sections by shifting all the predicted values. 
Comparisons between different models have to be done by using the same 
value of the axial mass.

The total quasi elastic neutrino and antineutrino cross sections for the three
flavours are presented in fig.\ref{figallmu} where the dashed lines show the 
FG results and the full lines have been obtained with our folding model.
As expected, the effect of the FSI is analogous for all the cases, 
since it depends on the nuclear structure and not on the details
of the interaction.

In Fig.\ref{figexp} we compare the $\nu_{\mu}$-nucleus cross section data 
\cite{kit83,bel85,bar77,bru90,bak82,vov79,far84,all90} with the 
FG result \cite{lip95} and with that of our model including FSI.
The contributions of the nucleon resonances and the deep inelastic scattering
to the total cross section (calculated in the framework of the FG model) are also
shown. The $15\%$ lowering of the quasi elastic peak, due to Final State 
Interactions is clearly visible. 

Finally a crude estimate of the effect of the lowering of the quasi-elastic peak
on the total upward muon flux, produced by atmospheric neutrinos, has also 
been done. Above a $\mu$ energy threshold of 100 MeV, the integral muon 
flux is reduced by $\sim 5 \%$.

\section{Summary and conclusions}
\label{conclu}
The $\nu$-nucleus scattering in the quasi elastic region leaves the
nucleus in a highly excited state which decays mainly by nucleon
emission. This decay is not properly treated in MF nuclear models,
where the emitted nucleon does not interact with the residual nucleus. 
We have presented a phenomenological model to correct the
MF cross sections for this effect.
The re-interaction model has been applied to FG differential cross
sections. It could also be applied to   
more sophisticated MF models. However, for quasi elastic scattering, 
our calculation takes into account, in a simple way, 
the main physical ingredients since in this energy region collective 
nuclear excitations and finite size effects are not important 
\cite{co88,bub91,eng93,ama94}. 
 
We have shown that the re-interaction spreads the cross section
strength at both higher and lower energies, with respect to the 
pure MF result. This produces a general lowering effect
of about $15\%$ in the cross section.

\vskip 1.cm
\noindent
{\bf Aknowledgement} \\
We whish to thank P. Lipari and G. Battistoni for useful discussions.

\vskip 1.cm
\section{Appendix}
In this appendix we briefly recall the link between response function, 
Green's function and cross sections. 
The commonly used approximations adopted in the derivation of 
the neutrino or antineutrino cross section on nucleus are:
first order pertubation theory, impulse
approximation and ultra-relativistic limit (i.e. the terms containig
the leptons rest masses are neglected).
With these assumptions the cross section 
can be written, in the lab frame, as \cite{wal75}:
\begin{eqnarray}
\displaystyle{ \frac{d^2 \sigma}{d \Omega_l d E_l}} &=&
\frac{g^2 E^2_l}{2\pi^2} \frac{4 \pi}{2J_i+1}
\Big\{
cos^2 \frac{\theta}{2} 
\Big[ \sum_{J=0}^{\infty} 
|<J_f|| {\cal M}_J - \frac{\omega}{|{\bf q}|}
{\cal L}_J ||J_i>|^2 
\Big] \nonumber \\
&~&
+\Big(\frac{q^2}{2{\bf q}^2}cos^2 \frac{\theta}{2}
+sin^2 \frac{\theta}{2} 
\Big) \nonumber \\
&~&
\times \Big[  \sum_{J=1}^{\infty}  
\Big(|<J_f||{\cal J}^{el}_J ||J_i>|^2 + 
|<J_f||{\cal J}^{mag}_J ||J_i>|^2 \Big)
\Big] \nonumber \\
&~&
\mp sin\frac{\theta}{2} \frac{1} {|{\bf q}|}
\Big(q^2 cos^2 \frac{\theta}{2} + {\bf q}^2 sin^2 \frac{\theta}{2} 
\Big)^{\frac{1}{2}} \nonumber \\
&~&
\Big[\sum_{J=1}^{\infty} 2 Re 
<J_f||{\cal J}^{mag}_J ||J_i><J_f||{\cal J}^{el}_J ||J_i>^*
\Big]
\Big\} \,\, ,
\label{fullcross}
\end{eqnarray}
where the $\mp$ should be used  for neutrino or
antineutrino scattering respectively. 
In the above expression we have indicated with $g$ the universal weak
coupling constant, with $E_l$ the energy of the emitted lepton,
with $q \equiv (\omega, {\bf q})$ the four-momentum
transfer, with $\theta$ the angle between incoming and outgoing
leptons, with $|J_i>$ and $|J_f>$ the initial and final states of the
nuclear system characterized by their total angular momenta $J$.
The quantities 
${\cal M}_J$, ${\cal L}_J$, ${\cal J}^{mag}_J$, and ${\cal J}^{el}_J$ 
are multipole expanded operators obtained by separating
the hadronic tensor in charge, longitudinal and transverse electric
and magnetic operators. As discussed above, these operators are
expressed in terms of one--body operators within the nuclear
many--body Hilbert space. 

In the expression (\ref{fullcross}) of the cross section
the leptonic variables are separated from the hadronic ones.
The information about the hadronic part is fully contained in the 
reduced matrix elements of the transition amplitude from the nuclear
ground state $|J_i>$ to the excited state  $|J_f>$.
In eq. (\ref{fullcross}) the sums over all the possible values of
$J$ have been included since we are interested in an energy range 
above the nucleon emission
threshold.  The nuclear 
states form a
complete basis since they are eigenstates of the nuclear hamiltonian $H$,
therefore it is possible to relate the transition
matrix elements to the linear response function as it is
usually defined in many--body theories \cite{kad62}:
\begin{eqnarray}
S (|{\bf q}| , \omega) 
&=& \sum_f <f|{\cal O}({\bf q})|i> 
         <f|{\cal O}({\bf q})|i>^\dagger
         \delta(E_f-\omega) \\
& =& \displaystyle{\frac{1}{\pi}} 
Im <i| {\cal O}^\dagger({\bf q})   
G( \omega ) 
{\cal O} ({\bf q})|i>.
\label{resdef}
\end{eqnarray}
$G( \omega )$ is the many--body Green function: 
\begin{equation}
G (\omega ) = 
\frac{1}{H- \omega - i \eta} - \frac{1}{H+ \omega + i \eta},
\label{fullgreen}
\end{equation}
where we have indicated with $H$ the full many-body hamiltonian.
In the above equations $|i>$ and $|f>$ indicates the initial and final
nuclear states, $E_f$ the eigenvalue of $|f>$, 
and ${\cal O} ({\bf q})$ a generic
many--body operator.

We are interested in nuclear excitation energies
whose values are well above the nucleon emission threshold i.e.
in reactions disintegrating the nucleus. Generally the
nuclear final state can be very complicated, for example, it can be
composed by a set of unbound nucleons or various nucleon clusters.
We make the simplifying assumption
that the nuclear final state is formed by one particle in the
continuum and a residual nucleus composed by $A-1$ nucleons. 
Even within this assumption the evaluation of the nuclear transition
amplitudes, or of the nuclear responses, is rather difficult if one
attempts to solve the nuclear many--body problem using the full
interacting hamiltonian $H$. 

A successful model used to obtain a simplified solution of
the full problem is the MF  model, where the
nucleons are supposed to move in an average potential independently
from each other. This means that in the above equations the
hamiltonian $H$ is substituted with a MF hamiltonian $H_0$
which is formed by a sum of single particle hamiltonians
whose eigenvectors form a basis of single particle states. 
           
\vskip .5cm


\newpage

\newpage

\newpage

\newpage

\newpage 

\end{document}